\font\fontA=cmss17 
\def\be{\begin{equation}}
\def\ee{\end{equation}}
\def\bea{\begin{eqnarray}}
\def\eea{\end{eqnarray}}

\def\nn{\nonumber}
\documentstyle[12pt]{article}
\def\pph{
\rightline{\today}
\rightline{RITSUMEI-PP-11}
\rightline{OCHA-PP-83}
\rightline{WU-AP/62/96}
}
\pph
\begin{document}
\par
\rightline{}
\begin{center}
{\fontA 
Late-Time Mild Inflation}\\[4mm]
{--- a possible solution of dilemma: cosmic age and the Hubble parameter ---}
\par
\vskip1truecm
{\sc Takeshi~FUKUYAMA}$^1$, {\sc Mikiko~HATAKEYAMA}$^2$, 
{\sc Masae~MIYOSHI}$^1$ \\[2mm]
{\sc Masahiro~MORIKAWA}$^2$ and,
{\sc Akika~NAKAMICHI}$^3$
\vskip1truecm
{\it Department of Physics, Ritsumeikan University$^1$ \\[2mm]
Kusatsu Shiga 525-77 JAPAN }
\vskip1truecm
{\it Department of Physics, Ochanomizu University$^2$ \\[2mm]
1-1, Otsuka 2, Bunkyo-ku, Tokyo 112 JAPAN }
\vskip1truecm
{\it Department of Physics, Waseda University$^3$ \\[2mm]
Ohkubo Shinjuku-ku, Tokyo 169 JAPAN }
\par
\vfill
\end{center}
\noindent{\bf Abstract}
\\[5mm]
We explore the cosmological model in which a late-time mild inflation is realized after the star formation epoch.    
Non-vanishing curvature coupling of a classical boson field yields this mild inflation without a cosmological constant.  
Accordingly the lifetime of the present Universe is remarkably increased in our model. 
Thus we show that the present observed high value of the Hubble parameter $H_0 \approx 70-80{\rm km/sec/Mpc}$ is compatible with the age of the oldest stars $14{\rm Gyr}$ without introducing the cosmological constant or the open Universe model.  
Moreover in our model, the local Hubble parameter becomes larger than the global one.  
Thus we show that the present observed local Hubble parameter measured by using the Cepheid variables is compatible with the global Hubble parameter measured by using the Sunyaev-Zeldovich effect.   
Furthermore we examine several aspects of our model:  
a)  The energy conditions in our model are violated.  We examine the consequences of these violations. 
b)  There is a natural evolution of the effective gravitational ``constant'' in high redshift region.  This yields drastic change of the stellar luminosity through the constructive equations of a star.  
We point out that a distant galaxy becomes much dimmer by this effect. 
c)  This varying Gravitational ``constant'' affects the cosmic expansion speed and the nucleosynthesis process in the early Universe.  
We point our that this effect constrains the parameters of our model though the fine tuning is always possible.  \\
\def\subjecthead{
\noindent
{\it Subject headings} : Cosmology : theory --- Hubble constant and the age of the Universe --- 
\hspace*{3.5cm}methods : numerical calculations\\
\noindent{\bf pacs number: }
}
\newpage
\section{Introduction}

\par
Recent many cosmological observations based on the method of Cepheid variables have consistently revealed the value of the present expansion rate of the Universe as $H_0=70-80 {\rm km/sec/Mpc}$ \cite{pierce94}\cite{freedman94}\cite{mould95}.  
These values are appreciably higher than what theoretical cosmologists expected before, $H_0 \approx 50$, mainly from the age analysis of the Universe.  
Actually the most simple cosmological model with no cosmological constant and no spatial curvature, $H_0=70-80 {\rm km/sec/Mpc}$ infers the age of the Universe as $8.3-9.4{\rm Gyr}$.  
This is an apparent conflict with the age of the oldest stars in the globular clusters, $14{\rm Gyr}$~\cite{walker92}.  
Stars should be younger than the Universe!  This is the {\it age problem of the Universe}.  

\par
The problem is not restricted to this.  
If we look into the cosmological observations on the expansion rate in detail, we realize, besides the age problem, that  there is an apparent discrepancy.  The local ($z \approx 0.001-0.004$) Hubble parameter is consistently larger than the global ($z \approx 0.17-0.18$) one.  
The local observations ~\cite{pierce94}\cite{freedman94}\cite{mould95} are mainly based on the luminosity-periodicity relation of Cepheid variable stars.  
They find Cepheid variable stars in M81, M100 and NGC4571, nearby galaxies of redshift $z \approx 0.001-0.004$.  The estimated Hubble parameter is $H_0=80-90 {\rm km/sec/Mpc}$.  On the other hand the global observations ~\cite{jones93}\cite{birkinshaw94} are mainly based on the Sunyaev-Zeldovich effect~\cite{sunyaev72}.   
They observe distant clusters of galaxies Abel2218 and Abel665 with redshift $z \approx 0.17-0.18$ and measured the temperature distortion induced by the inverse Compton scattering of 3K cosmic background radiation by hot gas around the cluster.  The estimated Hubble parameter is $H_0=50 {\rm km/sec/Mpc}$.  
If these observations are true and the local expansion rate is actually larger than the global one, then we have to reconsider the present standard homogeneous Universe model.  
Let us call this the {\it local-global $H_0$ discrepancy problem}\footnote{ Of course we have to wait for much decisive observations in the future.  
There may be any unknown systematic bias effects in the Sunyaev-Zeldovich method.  
There appear further complication if we consider another class of measurements of $H_0$ based on the supernovae of type I at maximum B light.  
It systematically shows the low value for the expansion rate $H_0=50 {\rm km/sec/Mpc}$ independent of the distance $0.004 < z <0.2$\cite{sandage82}. 
In this paper we do not consider these complications.  }. 

\par
Well known solutions for the first age problem are a) to consider an open Universe (low density $\rho$) or b) to introduce the positive cosmological constant ($\Lambda$).  This is manifest in the space-space component of the Einstein equation for the scale factor $a(t)$ in the homogeneous and isotropic matter dominated universe, 
\be
{\ddot a \over a}=-g(\rho+3P)+{\Lambda \over 3}c^2, 
\label{sseinstein}
\ee
where $g \equiv 4\pi G/3$.  
In the case a), the maximum lifetime of the Universe is $t_0= H_0^{-1}$.  
In the case b), the maximum lifetime is infinite if we fine tune the parameters: 
$\Lambda \rightarrow \Lambda_{\rm cr}\equiv (4/9)H_0^2a_0^2\Omega_0$ where 
$\Omega \equiv \rho/\rho_{\rm cr}, ~~\rho_{\rm cr} \equiv H_0^2/(2g)$.  This model is called as the Lema\^{i}tre Universe \cite{eddington30}\cite{lemaitre31}. 

\par
For the second local-global discrepancy problem, there are individual extensions of the above solutions a) and b).  
For the solution a), the authors of Ref. \cite{turner92}\cite{tomita95}\cite{tomita96} consider an inhomogeneous cosmology in which the local Universe is underdence and the local expansion rate is higher than the global average.  
Authors of \cite{nakamura95} try to constrain the maximum age of the inhomogeneous Universe based on the argument of Ref. \cite{landau}.  
It would be difficult to obtain an appropriate shape and configuration of the void around us naturally; it would require the anthropological principle.  
On the other hand for the solution b), the expansion rate does reduce in past.  However the reduction rate seems to be very small.  The appreciable reduction of $H_0$ is expected only when $z \geq 1$.   

\par
We would like to improve unsatisfactory points of the previous models and would like to propose a new model.  
We now consider the model which satisfies the following requirements:  
\begin{enumerate}
\item The Universe has a long lifetime which is compatible with the age of the 
oldest stars (age problem).  
\item Local ($z \approx 0.001-0.004$) Hubble parameter is larger than the global ($z \approx 0.17-0.18$) one:  
$H_0$(local)$>$$H_0$(global) (local-global discrepancy problem).  
\item We do not consider a dilute Universe.  We fix $\Omega \equiv 
\rho/\rho_{\rm cr}=1$ (flat Universe).  
\item As in the Lema\^{i}tre Universe model, we allow some amount of fine 
tuning for solving the age problem.  
\end{enumerate}

\par
We easily realize from Eq.(\ref{sseinstein}) that large {\it negative pressure} in some epoch tends to increase the acceleration ($\ddot a$) of the scale factor and extends the cosmic lifetime in the similar way as the small density $\rho$ and the positive cosmological constant $\Lambda$.  
This is similar to the idea of cosmological inflation in the early Universe \cite{guth81}\cite{sato81} .  
However for our purpose to solve the age problem, the inflation must be in the {\it late-time} of the Universe definitely after the star formation epoch.  
Moreover the original scenario of inflation in the early stage does not contribute to the cosmic age because the scenario has too high energy scale such as $10^{15} {\rm GeV}$.    

\par
Therefore we explore the late-time inflationary model after the star formation epoch and study the observational consequences of this model.  
Apparently this late-time inflation should be {\it mild} in the sense that the total expansion of the Universe during this inflation should be small compared with that in the original inflationary model $ \approx e^{60}$.  
In order to yield this late-time inflation, we introduce a hypothetical {\it scalar field with an ultra-light mass and a strong curvature coupling}.  
This type of model has been widely used so far in various contexts in cosmology \cite{la89},\cite{fakir90}, and \cite{morikawa91}.  
We mainly follow these phenomenological arguments in this paper\footnote{ However in some stage of future researches this kind of scalar field should be identified.  See the discussions in the last section.  }.
Our late-time inflationary model turns out to solve the local-global $H_0$ discrepancy problem as well as the age problem.  
This is because our model is highly non-linear due to the curvature coupling and the cosmic expansion is not monotonic.  
The mass of the scalar filed yields oscillatory cosmic expansion rate which can appreciably change within the redshift interval 0.01.   

\par
Originally Fakir et al. In Ref.\cite{fakir90} pointed out the possibility of inflation in the model of scalar field with negative\footnote{Their $\xi$ is minus of our $\xi$.} curvature coupling $\xi$ with an appropriate potential.    
They studied the primordial density fluctuations in their inflationary model.  
Here we would like to apply the similar model to the late-time cosmology after the star formation epoch.  
The negative signature of $\xi$ is essential for the inflation and the elongation of the cosmic age.  
This will be explained in detail in subsequent sections.   
Therefore, for example, our previous model ~\cite{morikawa91} where we took positive $\xi$ did not show any inflation nor elongation of the cosmic age.  

\par
Moreover in our model the effective gravitational constant $G_{\rm eff}$ changes in time because the scalar field directly couples with the scalar curvature.  
This may cause the intrinsic change of the stellar luminosity which is known to be very sensitive to the strength of gravity.  
Actually in the purely inflating phase in our model, the effective gravitational constant becomes half of the present value and the stellar luminosity becomes about six magnitude dimmer than the normal stars now.  

\par
The construction of this paper is as follows:  
In Section 2, we introduce a model of scalar field with a curvature coupling and show the existence of the mild inflation phase.   
In Section 3, observational properties of this mild Inflation model is explored.  
We show the cosmic age elongation and the non-linear effect that the local expansion rate becomes higher than the global one.  
In Section 4, we examine our mild inflationary scenario from wider point of view.  
We first examine the energy conditions and show that they are violated. Implications of this fact are given.   
Then we point out that the effective gravitational constant reduces toward past and the possible reduction of the intrinsic luminosity of stars.    
We also mention the effect of varying $G_{\rm eff}$ on the nucleosynthesis.  
In the last Section 5, we summarize our present work and discuss on our future directions of research. 
 
\section{The Model of Mild Inflation}
Let us begin with the following action 
\be
S=\int d^4 x \sqrt{-g}
\left[ {1 \over 2}g^{\mu\nu} \phi_{,\mu}\phi_{,\nu} 
- {1 \over 2}(m^2-\xi R)\phi^2-{R \over 16\pi G}+{\cal L}_m
\right].  
\label{action}
\ee
The scalar field $\phi$ with mass $m$ couples with the scalar curvature $R$ with strength $\xi$ which is taken to be negative.  
The last term on the right hand side of the above equation ${\cal L}_m$ represents the ordinary matter whose energy density is given by $\rho_m$.  
We assume the isotropic and homogeneous Universe, the Friedman Universe, with no spatial curvature.  
Then the line element is given by 
\be
ds^2=dt^2-a^2(t)[d\chi^2+\chi^2d\Omega^2].  
\label{lineelement}
\ee
In this Universe, the spatial gradient term of the scalar field tends to reduce in the course of cosmic expansion.  
Therefore we only consider the spatially uniform configuration for the scalar field from the beginning.  
The energy-momentum tensor is defined as $T_{\mu\nu}\equiv 2 (-det~g_{\mu\nu})^{-1/2}~\delta S /\delta g^{\mu\nu}$ and the corresponding density and pressure are given by 
${\rm diag}T_{\mu}^{\nu}=(\rho, p,p,p)$ with 
\bea
\rho&=&{1 \over 2}(\dot \phi^2 + m \phi^2)+3 \xi \phi^2 (H^2 +{k \over a^2})+6 H \xi \phi\dot\phi + \rho_m, 
\nonumber \\
p&=&{1 \over 2}\dot \phi^2 -{1 \over 2}m^2 \phi^2 + \xi\phi^2(-3 H^2 -{k \over a^2}-2 \dot H)-4 H \xi \phi \dot\phi -\xi (\phi^2\ddot),
\label{rhop}
\eea
where $k$ is the curvature constant which we eventually set 0.  
$H$ is the Hubble parameter $H \equiv \dot a/a$.  
Then the equation of motion for the scalar field becomes  
\be
\ddot{\phi}+3H\dot{\phi}+(m^{2}+6\xi(\dot{H} +2H^{2}+ka^{-2}))\phi=0.
\label{eq-scalar}
\ee
Time-time component of the Einstein equation becomes
\be
H^2+ka^{-2}=g(\dot\phi^2+m^2\phi^2+6\xi\phi^2(H^2+ka^{-2})
               +12\xi H\phi\dot\phi+2\rho_{m}), 
\label{time-time}
\ee
and the space-space component becomes
\bea
& &\dot H (6\xi\phi^2(1-6\xi)-g^{-1}) 
=3\dot\phi^2 +6\xi(4H\phi\dot\phi-\dot\phi^2+(m^2+12\xi H^2)\phi^2)
\nonumber \\
&&\mbox{\hspace*{5cm}}
+ka^{-2}(6\xi\phi^2+{1\over 2}g^{-1})+3\rho_{m},
\label{space-space}
\eea
where $g=4\pi G/3$.  

\par
This set of equations admits a special solution if the parameter $\xi$ is 
negative, $k=0$, and $\rho_{m}=0$:
\bea
\dot\phi&=&0, ~~ \dot H=0,  \nonumber \\
H_*^2&=&m^2/(12|\xi|), ~~
\phi_*^2=1/(6|\xi|g).  
\label{static-solution}
\eea
This is nothing but the de Sitter space-time the exponentially inflating Universe.  
In this solution, the effective mass of the scalar field ($m^{2}+6\xi(\dot{H} +2H^{2}+ka^{-2})$) vanishes and the expansion rate $H$ is frozen.  
Therefore the evolution of the scale factor becomes convex 
$\ddot a/a=H_*^2>0$ and the age of the universe is infinitely elongated.  
It should be emphasized that this solution is possible only for negative curvature coupling $\xi<0$ \footnote{Another choice $m^2 <0, ~~\xi>0$ also admits this solution though unrealistic.  }.  
If positive, the age of the universe would have been {\it reduced}.   
\par
This solution should be compared with the extreme Lema\^{i}tre universe, the Einstein static universe, in which the scale factor is frozen.  
Also in this case, the age of the Universe is infinitely elongated.  
However this Lema\^{i}tre universe is unstable; any small amount of perturbation forces the Universe collapse or re-expand forever.  
Similar but much milder instability exists in our model.  
A small deviation from this $H$-frozen solution is shown to be unstable.  Let us consider small deviations $x$ and $y$:  
\be
\phi=\phi_*(1 + x), ~~ H=H_* (1+ y).  
\label{deviation}
\ee
Then the linearized equations of motion for the deviations are 
\be
\left( {\matrix{x\cr
y\cr
}} \right)'
=
\left( {\matrix{1&-2\cr
{2 \over 1+3|\xi|}&-4\cr
}} \right)
\left( {\matrix{x\cr
y\cr
}} \right), 
\label{lineareq}
\ee
where the prime denotes the derivative with respect to the time measured by the 
frozen Hubble parameter $H_* t$.  
The above matrix has one positive eigenvalue around $1$ and one negative eigenvalue around $-4$ for $|\xi|\gg 1$.  
Therefore the $H$-frozen solution is unstable and the inflation eventually terminates.  
However the scalar field is globally oscillating due to the finite mass $m$ and the field configuration returns closely to the original configuration.  
Then the Universe reenters the inflationary phase.  
After some amount of inflation the configuration deviates from the H-frozen phase again.  
This behavior repeats several times during the cosmic expansion.  
Accordingly the Universe shows {\it piece-wise inflationary expansion} repeating the {\it H-frozen phase}.  
This is the essence of the mild inflationary model.  

\par
In order to have a rough idea for the mild inflation, we numerically solve the above set of equations.  
A typical example is shown in Fig. 1 and Fig. 2.  
\vskip1truecm
\centerline{Fig. 1. }
\vskip1truecm

\vskip1truecm
\centerline{Fig. 2. }
\vskip1truecm
We took our parameters as follows: 
\be
m=100 H_0, ~~
k=0, ~~
\xi=-80, ~~
\Omega_{\rm matter }=0.01, 
\label{parameters}
\ee
The above choice of mass $m$ ( $\approx 10^{-31}{\rm eV}$) yields several H-frozen phases until present time.  
If the mass were much larger than the above choice, then the piece-wise inflation would become less prominent and the energy density starts to behave as the ordinary dust.  
The above choice of curvature coupling $\xi$ yields appropriate strength in the oscillation of the Hubble expansion rate.  
If $\xi$ were much smaller than the above choice, then the inflation strength would become weaker.  
Spatially flatness, $k=0$, is assumed for simplicity and the density parameter of matter $\Omega_m$ is set to be the lowest possible value from typical  observations\footnote{ If we increase $\Omega_m$, the inflationary period and the age of the Universe would be reduced.  }.    
In this paper, we would like to propose a new type of cosmology and would like to show qualitative characteristics of the model.  
Therefore we will not try the best fit of our model with observations.  
In this sense the above choice of parameters are somewhat tentative though we will not change the choice in this paper.  
We set the ``final'' conditions at the present time $t_0$ as
\be
\phi(t_0)=0.0023 \sqrt{4\pi G/3}.  
\label{finalconditions}
\ee
This value is also somewhat arbitrary\footnote{As we will see in later in Fig. 3, the value 0.0023 for $\phi(t_0)$ is chosen in favor of longer age of the Universe.  }.    
Setting the value $\phi(t_0)$ and the present value of the expansion rate $H_0$ automatically fixes the value $\dot\phi(t_0)$ through the constraint equation Eq.(\ref{time-time}).  
We notice in Fig. 1 that the piece-wise inflation is realized in the scale factor.  
The oscillatory nature of the scalar field becomes manifest in Fig. 2. 
The scale factor vanishes at $t=-1.45802 H_0^{-1}$, with which we identify the point of big bang.  

\par
Several comments on this $H$- frozen phase are in order.  
1) 
The condition $\xi<0$ is inevitable for the existence of this phase.  
As we can see in the numerical calculation in Fig. 1 that the curve of the scale factor vs. the cosmic time becomes piece-wise convex downward for $\xi<0$.  
On the other hand if $\xi>0$, the graph would become piece-wise concave.  
The former convex case the cosmic age is increased and the latter concave case it is reduced.  
2) 
This type of inflation induced by the negative curvature coupling was studied in the paper \cite{fakir90} and mainly applied to the density perturbations in the early Universe.  
We are applying the similar model to the directly observable universe.    
In Ref. \cite{fakir90}, the existence of the self coupling of the scalar field and very large curvature coupling $|\xi|$ were essential for their inflation.  
Contrary to their case, we do not need the self coupling term and $|\xi|$ can be smaller than their case for our type of inflation.  
3) 
In Ref. \cite{futamase89}, the absolute value of the parameter $\xi$ must be much smaller than one for the inflation to occur if there is no self coupling term for the scalar field.  
However in the present case, this analysis does not apply.  
This is because we are now studying the transient oscillatory model of inflation which was not in the scope of Ref. \cite{futamase89}.  
Actually the pure inflationary solution Eq.(\ref{static-solution}) is not an attractor in our model and the permanent inflation is not realized.  
However the piece-wise inflation is realized due to the several repetitions of the scalar field to the phase Eq.(\ref{static-solution}).  
\section{Observational properties of the Mild Inflation}
We now study the observational properties of the mild inflationary model of Universe.  
\subsection{age of the present Universe}
Apparently the existence of the inflationary phase extends the age of the Universe.  
This is because the curve of the scale factor vs. cosmic time becomes convex downward as we see in Fig. 1.
However actually, the late-time inflation cannot continue forever.  
If it were the case, the cosmic matter baryonic density would be thoroughly diluted away; manifest conflict with the observations.  
According to the recent observations the baryonic matter should occupy one to ten percent of the total matter density of the Universe. \cite{boerner93}  
This fact most severely constrains the total amount of inflation.
Baryonic matter density would be diluted, in the pure inflationary phase, by the factor $(a_f/a_i)^3$ where $a_f$ and $a_i$ are the scale factors at the initial stage and final stage of the inflation phase, respectively.   
Our late-time inflation is not a pure inflation but piece-wise inflation.  
Therefore at each inflationary phase, some amount of the baryonic matter density is diluted away and the total accumulation of the loss determines the present baryonic matter density.  
Stronger inflation much elongates cosmic age but too much inflation makes the Universe empty.  
We need compromise and took $\Omega_{\rm matter}=0.01$ in our numerical calculations.  

\par
In our numerical calculations in the previous section, we set the parameters as Eq.(\ref{parameters}) Eq.(\ref{finalconditions}).  
We obtained the present age of the Universe $t_0=-1.45802 H_0^{-1}$ by identifying the big bang at the point when the scale factor vanishes (Fig.1).  
Then the age of the present Universe is read as $17.86 {\rm Gyr}$ if we adopt $H_0=80{\rm km/sec/Mpc}$.  
This is not a very special value.  
The age of the Universe is naturally elongated.  
We numerically checked this by varying the ``final'' condition $\phi_0$ fixing all the other parameters. 
We show the result in Fig. 3.  

\vskip1truecm
\centerline{Fig. 3. }
\vskip1truecm

\noindent
According to Fig. 3, we obtain the cosmic age larger than $17 {\rm Gyr}$ for the parameter range $\phi(t_0)/\sqrt{4\pi G/3}=0.0015-0.006$.  
This variation of $\phi_0$ is equivalent to the variation of the identification of the oscillation phase at the present time.  
If the phase is chosen so that the present expansion rate becomes local maximum, then it corresponds to the local maximum of the cosmic age.  
Therefore each peak in Fig. 3 corresponds to the local maximum of the present expansion rate.  
The total number of oscillations or the total number of H-frozen phases is different from peak to peak in Fig. 3.    
At the peak $\phi_0=0.0023$ (which we have chosen before in the previous section), the number of H-frozen phase becomes maximum.   
The number of H-frozen phase for each peak is denoted in Fig. 3.  
Note that larger number of H-frozen phase yields larger cosmic age.  

\subsection{local and global Hubble parameters}
\par
Now we turn our attention to the second problem the local-global discrepancy of $H_0$.   
In our model, the expansion rate of the Universe is oscillating around the globally reducing component.  
Therefore, depending on the phase of the oscillation, the Hubble parameter averagingly smaller in the past than the value at present.  
It's typical reduction time scale is given by the oscillation period.  
For example in the previous numerical calculations, the change of the Hubble parameter in the small redshift region is shown in Fig. 4.  
\vskip1truecm
\centerline{Fig. 4. }
\vskip1truecm
The dashed line in this graph represents the expansion rate $H \equiv \dot a(t)/a(t)$ versus redshift $z$.  
However this is the bare Hubble parameter and is different from the observable Hubble parameter $H_{\rm obs}$.  
The observable Hubble parameter is usually obtained either from 
a)  the distance measured through the luminosity of a star or 
b)  the distance measured from the angular diameter of an X-ray cluster.  
These distances are translated into the coordinate distance $\chi$ in Eq.(\ref{lineelement}).  
The coordinate distance $\chi_1$ to an object which emits light at the cosmic time $t_1$ and shows redshift z is simply given by integrating $d\chi$ in Eq.(\ref{lineelement}) along the light path ($ds^2=0$) aided by Eq.(\ref{time-time}).  
In the simplest standard cosmology $\Omega=1, ~ \Lambda=0$, it becomes 
\be
\chi_1=\int_{a_1}^{a_0}{dt \over a(t)}
={2 \over H_0 a_0} (1-(1+z)^{-1/2}).  
\label{cdistance}
\ee 
Therefore the observable Hubble parameter is obtained from this relation as  
\be
H_0={2 \over \chi_1 a_0} (1-(1+z)^{-1/2}) \equiv H_{\rm obs}.  
\label{hubble}
\ee
This observable Hubble parameter is plotted in Fig. 4 with the solid  line.  
This observable parameter behaves much milder than the bare Hubble parameter because it is an integral of the inverse of the scale factor which is already an integral of the oscillating quantity $H$.  
It is obvious from Fig. 4 that $H_{\rm obs}$ rapidly reduces in past and stays almost constant with low value.  
Already at $z=0.04$ the Hubble parameter $H_{\rm obs}$ reduces almost 60 percent of the present value.  
This is the characteristic of the recent observations on the Hubble parameter mentioned in the introduction.  
Typical observed values $H_{\rm obs} \approx 70-80 {\rm km/sec/Mpc}$ at $z \approx 0.001-0.004$ and 
$H_{\rm obs} \approx 50 {\rm km/sec/Mpc}$ at $z \approx 0.17-0.18$ are qualitatively consistent with our results of numerical calculations.    

\par
For a comparison, we also plotted the Hubble parameter in the simplest standard model by the thin solid line.  
The observed Hubble parameter $H_{\rm obs}$ defined above becomes constant in this case.  

\par
The reduction time scale is basically determined by the oscillation period of the expansion rate which is controlled by the mass parameter $m$ of the scalar field.  
The strength of reduction is basically determined from the non-linearity of the set of evolution equations Eqs.(\ref{eq-scalar}),  (\ref{time-time}), and (\ref{space-space}) controlled by the parameter $\xi$.  
Though we do not try to fine tune these parameters to fit to the observations in this paper, these parameters may become important in the near future when many reliable observations are available.  

\par
The reduction of $H$ or $H_{\rm obs}$ toward the past is closely connected with the elongation of the age of the Universe.  
Suppose we vary the present phase of the oscillation, or equivalently the ``final'' condition $\phi(t_0)$.  
If the phase is chosen so that we observe the most decreasing Hubble parameter toward the past, then this is the most favorable situation for the age problem.  
The age of the Universe in this case becomes a local maximum in the variation of the phase.  
This is because, in this situation, we are now located in the phase of $H$ local maximum whose value should be fixed by observations.  
Therefore the global average of the Hubble parameter becomes minimum and the cosmic age becomes local maximum.  
This specially ``lucky'' phases correspond to several peaks in Fig. 3.  
This situation seems to be very similar in appearance to the model of inhomogeneous Universe in  Refs. \cite{turner92}\cite{tomita95}\cite{tomita96}.  
However, unlike the inhomogeneous model, we do not have to rely upon the anthropological principle in order to set our location just in the center of a void.   
In our model, any observer at present time $t_0$ observe the virtual void structure with the observer located just on the center.   
This is because the change of $H$ in redshift $z$ is caused by a temporal structure in the homogeneous Universe and is not caused by real inhomogeneity .  

\par
For the comparison with the simplest standard cosmology, we  have shown in Fig. 5. the m-M and Log(z) relation (Hubble diagram ), where $m-M$ is the distance modulus the difference of the apparent luminosity and the absolute luminosity of galaxies.  
We have chosen the present Hubble parameter $H_0=80 {\rm km/sec/Mpc}$ and therefore
\bea
m-M&=&5 \log_{10}((1+z)\chi)-5(\log_{10}(H_0 {\rm pc}))-5 \\
&=&\log_{10}((1+z)\chi)+42.87.  
\label{hubblelti}
\eea
The thick solid line in this graph represents this relation which is expected in our model.  

\vskip1truecm
\centerline{Fig. 5. }
\vskip1truecm
\noindent
In the simplest standard model with $\Omega=1, ~\Lambda=0$, the relation is given by  
\bea
m-M&=&5 \log_{10}(2 (1+z) (1-{1 \over \sqrt{1+z}}))-5(\log_{10}(H_0 {\rm pc}))-5 \\
&=&\log_{10}((1+z)(1-{1 \over \sqrt{1+z}}))+42.87.  
\label{hubblestd}
\eea
The thin solid line in Fig. 5 represents this relation.  
Since larger distance modulus $m-M$ means smaller $H_0$, it is obvious in Fig. 5 that $H_0$(local)$>$$H_0$(global) is actually realized.

\section{Examination of the model}
In this section, we examine other aspects of our late-time inflationary model.  
Because this model is a new proposal for the solution of cosmological problems, we need to examine it in a wide point of view.  
We simply examine the natural consequences of our model and do not try to justify our model\footnote{In particular, we do not further fine tune the  parameters of the model.  }.  
This is because we believe that there are many points to be developed in our model and therefore it is too early to judge the final validity of the model.  
Among various interesting aspects of this model we picked up three topics:
Violation of energy conditions, The luminosity biasing and Nucleosynthesis.  

\subsection{Hubble parameter and energy condition}
\par
The elongation of the cosmic age is closely related with the violation of energy conditions.  
We first examine these energy conditions in our model.  

\par
According to the argument in the reference \cite{nakamura95} ,  the volume expansion rate 
\be
H_v \equiv {1 \over 3} {1 \over \sqrt{\det \gamma}} {\partial \over
 \partial t} \sqrt{\det \gamma} , 
\label{volumeexpansionrate}
\ee
with $\gamma_{ij}$ being the spatial metric,  
satisfies the inequality
\be
H_{v}(t_0)t_0 \le 1
\label{limit}
\ee
under the following conditions  
\begin{enumerate}
\item[1)] Einstein's theory of gravity is correct.
\item[2)] Caustic(at which $H_v=\infty$) is identified with the Big Bang Singularity.
\item[3)] $H_v(t_0)>0$
\item[4)] The strong energy condition holds:
\be
R_{\mu \nu} V^{\mu} V^{\nu} \ge 0
\label{sec}
\ee
for any timelike vector $V^{\mu}$.
\end{enumerate}
The volume expansion rate $H_v$ reduces to the Hubble constant in the homogeneous and isotropic Universe.  
The inequality Eq.(\ref{limit}) is essentially exhausted in the Landau-Lifshitz's textbook \cite{landau}:
\be
{\partial \over \partial t} {1 \over H_v} \le 1.  
\label{landau}
\ee
In our late-time inflationary model, the first three conditions 1) 2) 3) hold but the last condition 4).
Let us investigate this point.  
The strong energy condition becomes
\be
\rho + p > 0, ~~{\rm and}~~\rho +3p > 0
\label{secrhop}
\ee
if we diagonalize the energy momentum tensor.
We can examine these conditions using the expression Eq.(\ref{rhop}).  
We numerically calculated $\rho$, $\rho+p$ and $\rho-|p|$ in Fig. 6a $\sim$ 6c.
\vskip1truecm
\centerline{Fig. 6}
\vskip1truecm
As we can see immediately the strong energy condition is apparently periodically violated.  
This is the reason why we could obtain the long lifetime of the Universe free from the constraint Eq.(\ref{limit}).  

\par
However we readily recognize that the weak energy condition and dominant energy condition are also violated as well as strong energy condition.  
The weak energy condition is the requirement that the energy density is non-negative for any observer.
That is, 
\be
T_{\mu \nu} V^{\mu} V^{\nu} \ge 0
\label{wec}
\ee
for any timelike vector $V^{\mu}$.
In our case, it reduces to
\be
\rho > 0, ~~{\rm and}~~\rho +p > 0.  
\label{wecrhop}
\ee
Our numerical calculation shows that our model violates the second inequality though the first inequality is satisfied.  
The dominant energy condition is the requirement that the energy flow $T_{\mu \nu} V^{\mu}$ is non-spacelike for any timelike vector $V^{\mu}$.
In our case, this reduces to
\be
\rho > 0, ~~{\rm and}~~\rho -|p| > 0.  
\label{decrhop}
\ee
Our numerical calculation shows that our model violates the second inequality.  
These violations seem to be a general feature of the model with the scalar field with curvature coupling.  
Actually even if we take other coupling constants $\xi$ such as +80 or +10, the weak and dominant energy conditions are violated.  
Therefore it is clear that this scalar field cannot couple with the ordinary observable matter.  
Because otherwise we would see the negative energy density and therefore the stability of matter is not guaranteed\footnote{If there exists observable matter with negative energy density, then this matter is highly unstable and therefore there will be a violent activity associated with this matter.}.

\subsection{Effective Gravitational Constant}
\par
The curvature coupling of the scalar field in our model yields the temporal change of the effective gravitational constant.  
From the action Eq.(\ref{action}), the effective gravitational constant is read as the inverse of the coefficient of the scalar curvature $R$.  
\be
G_{\rm eff}(t)={1 \over G^{-1}-8\pi \xi \phi(t)^2}.   
\label{effectiveg}
\ee
This parameter characterizes the gravitational interaction with time scale shorter than that of the scalar field.  
The latter is roughly determined by the inverse of the mass 
$m^{-1}=0.01 H_0^{-1}$ as we see from Eq.(\ref{parameters}).   
Because the parameter $\xi$ is taken to be negative in our model, $G_{\rm eff}(t)$ reduces on average if we go back in time.  
For example, $G_{\rm eff}(t)$ in the purely H-frozen phase exactly becomes one half of the present value.  
Actual temporal change of $G_{\rm eff}(t)$ is numerically calculated and shown in Fig. 7.

\vskip1truecm
\centerline{Fig. 7.}
\vskip1truecm
Within the redshift one, $G_{\rm eff}(t)$ changes up to five percent.  However it drastically changes at around redshift two and beyond.    

\par
On the other hand, the luminosity of a star is known to be very sensitive to the value of the gravitational constant.  
Now we examine this effect.  
The basic macroscopic equations for a spherically symmetric star are the equation of hydrostatic equilibrium, mass continuity, energy continuity, and radiative transport equations \cite{hansen94}:
\bea
{dp \over dr}&=&-{\rho G M(r) \over r^2} ,\\
{dM(r) \over dr}&=&4\pi r^2 \rho , \\
{dL(r) \over dr}&=&4\pi r^2 \rho \epsilon , \\
{dT(r) \over dr}&=&-{3 \kappa \rho L \over 16 \pi a c T^3 r^2} , 
\label{macroeqs}
\eea
where
\be
{ac \over 4} = {\pi^2 c k^4 \over 60 c^3 \hbar^3}
\ee
is the Stefan-Boltzmann constant.  
Here $r$ is the radial distance measured from the stellar center.
$\rho$, $L(r)$, and $T(r)$ are the mass density, the luminosity and the temperature at the position $r$, respectively.
$M(r)$ is the mass contained within a sphere of radius $r$.
These equations must be solved simultaneously with the microscopic equations, state equation, energy generation equation, and the equation for the opacity:
\bea
p&=&{\rho \over \mu m_H}kT, \\
\epsilon&=&\epsilon_0 X Z T^n, \\
\kappa_{bf}&=&\kappa_0 Z (1+X) \rho T^{-3.5},
\label{microeqs}
\eea
where $X$ and $Z$ are the hydrogen and metal mass fractions respectively. 
$\kappa_{bf}$ is the absorption (opacity) of a photon by a bound electron.
They form a complicated set of equations which can be exactly solved only by numerical calculations.  
Here in order to obtain a rough analytical estimate, we replace the differentiation such as $dp / dr$ by the ratio of averaged variables such as $p / R$.  
Then we obtain
\bea
R&=& b G^{(2n-15 )/(2n -1)} M^{(2n-9)/(2n-1)}, \\
T&=& \mu m (k b)^{-1} G^{14/(2n -1)} M^{8/(2n-1)}, \\
L&=& \epsilon_0 X Z ({\mu m \over k b})^n G^{14n/(2n -1)} M^{(10n-1)/(2n-1)}, \\
T_s&=& (\pi a c)^{-1/4} (\epsilon_0 X Z)^{1/4} {\mu m \over k}^{n/4}
b^{-(2+n)/4} G^{(5n+15 )/(4n-2)} \nn \\
&&M^{(6n+17)/(8n-4)}, 
\label{rough}
\eea
where
\be
b \equiv \left[  {3^3 \kappa_0 \epsilon_0 \over 4^4 \pi^3 a c 
}\right]^{2/(2n-1)}
\left[ \mu m \over k \right]^{(2n-15)/(2n-1)}
\left[  Z X (1+X) \right]^{2/(2n-1)}.  
\label{defab}
\ee
For example if the star is on the stage that the CNO-cycle is the dominant energy generation process, then $n=15$ and 
\bea
b&=&{\rm const.} [Z X (1+X)]^{0.069},\\
R&=&{\rm const.} [b G^{0.52} M^{0.72}],\\
T&=&{\rm const.} [b^{-1} G^{0.48} M^{0.28}],\\
L&=&{\rm const.} [X Z b^{-15} G^{7.2} M^{5.1}],\\
T_s&=&{\rm const.} [(XZ)^{1/4} b^{-4.3}G^{1.6} M^{0.92}].  
\label{n15}
\eea
If it is on the stage that the p-p chain is the dominant energy generation process, then $n=4$ and 
\bea
b&=&{\rm const.} [Z X (1+X)]^{0.29},\\
R&=&{\rm const.} [b G^{-1} M^{-0.14}],\\
T&=&{\rm const.} [b^{-1} G^{2} M^{1.14}],\\
L&=&{\rm const.} [b^{-4} G^{8} M^{5.57}],\\
T_s&=&{\rm const.}[(XZ)^{1/4} b^{-2} G^{2.5} M^{1.46}].  
\label{n20}
\eea
In both cases we notice large exponents 7.2 (for $n=15$) and 8 (for $n=4$) on $G$ for the stellar luminosity $L$.  

\par
This sensitive dependence of L on G yields a drastic results in our cosmological model.  
According to the above estimate, the stars in the redshift range from zero to 0.2, $G_{\rm eff}(t)/G_0$ is about 0.98 and the luminosity fluctuate only fifteen percent.  
On the other hand, the stars in the redshift range from two to five, $G_{\rm eff}(t)$ is about 0.8 on average and the luminosity fluctuate about 85 percent.  
Furthermore if the redshift exceeds about ten, the stellar luminosity reduces more than several hundreds times and actually such stars cannot be observed at all.  
It would therefore become very difficult to detect, in high redshift region, the ordinary stars and clusters of them shining by the nuclear reactions triggered by the self gravity.  
If we fined luminous object with high redshift, the emission process of it would be very different from the ordinary stars.  
Observationally, the most distant galaxy which is confirmed to have its luminosity from the collection of ordinary stars is located at around $z=3$\cite{cowie95}.  
Therefore it may be interesting to examine the possible change of galactic spectrum as a function of redshift, though actually the evolutionary effect makes the spectrum complicated.  
The above reduction of stellar luminosity at high redshift region may also cause the reduction of the number counts of galaxies in the past.   

\subsection{Nucleosynthesis}
We have previously explored mainly the directly observable epoch in the Universe i.e. the redshift less than about ten.  
However a naive extrapolation of our model into much early epoch may be interesting though there is no guarantee that the dynamics of the scalar field and its coupling to the scalar curvature $\xi$ are not changed from those at present.  
We consider the effect of $G_{\rm eff}(t)$ on the nucleosynthesis in this subsection.  

\par
In the early Universe the energy density is expected to be dominated by radiation and the scalar field contribution to the total energy density becomes less important.  
However since the scalar field couples with the scalar curvature, it strongly affects the effective gravitational constant even in the early Universe.  
Actually the asymptotic solution of Eq.(\ref{eq-scalar}) in the radiation dominated era ($H=1/(2t)$) becomes 
\be
\phi=c_1 +c_2 t^{-1/2}
\label{asymptotic}
\ee
with $c_1,c_2$ some constants.  
We can safely discard the growing mode toward the past ($c_2=0$) if we assume that the scalar field was finite in the past.  
Even in this case $G_{\rm eff}(t)$ in the epoch of nucleosynthesis might be generally different from the present value.  
Since the value of $G_{\rm eff}(t)$ determines the expansion rate at that time, it strongly affects the dynamics of nucleosynthesis and the primordial abundance of $^4$He.   
According to the arguments in Ref. \cite{accetta90}, the constraint is obtained as
\be
{|G_{\rm eff}(t_{\rm ns})-G_0| \over G_0} < 0.4,   
\label{Gconstraint}
\ee
where $t_{\rm ns}$ is the time of nucleosynthesis.  
Though it is always possible to fine tune the parameter $c_1$ so that the above constraint is satisfied, it goes beyond the scope of the present paper to argue what guarantees this tuning.  

\par
On the other hand we doubt that our scalar field with tiny mass ($10^{-31} {\rm eV}$) is the elementary field even at the epoch of nucleosynthesis (energy scale of MeV).  
If it decomposes into other elementary fields or its curvature coupling vanishes at the epoch of nucleosynthesis then $G_{\rm eff}(t_{\rm ns})$ is the same as $G_0$ and it does not affect the dynamics of nucleosynthesis.

\section{Conclusions and discussions} 
\par
We first pointed out two basic problems in the present cosmology: 
a)  The age problem and 
b)  the local-global $H_0$ discrepancy problem.  
In order to solve these problems we have introduced a cosmological model with scalar field which has negative curvature coupling constant $\xi$.  
In this model, we have obtained sufficient amount of elongation of the cosmic age:  $17.86 {\rm Gyr}$ if we adopt $H_0=80{\rm km/sec/Mpc}$ for an appropriate choice of parameters.
Moreover this model shows approximately forty percent reduction of $H_{\rm obs}$ within $z \approx 0.04$ and this may partially explain the low values of $H_{\rm obs}$ measured by the method of Sunyaev-Zeldovich effect.   

\par
In a simple open Universe model the cosmic age is also elongated however the Hubble parameter never reduces toward the past.  
In a model with cosmological constant the cosmic age is also elongated and the Hubble parameter can reduce toward the past however the reduction rate is extremely small (about one percent at redshift 0.04).  
In a model with inhomogeneous matter distribution, the rapid reduction of the Hubble parameter toward the past may be accounted for but the age problem still remains.  
Our late-time inflationary scenario will be the unique model which can account for both the rapid reduction of $H$ and the age problem.  

\par
These extreme properties are due to the piece-wise mild inflation realized in our model. 
For this mechanism, large negative $\xi$ has been essential.   
If we would have taken positive $\xi$, the cosmic age would have been shorten.  

\par
Then we have examined couple of consequences of the late-time inflationary scenario.  
They constrain the parameters and simultaneously indicate interesting developments of our model.  
Both of them should be continuously studied and will be reported soon.  
\begin{enumerate}
\item[1.]
Our model violates all energy conditions, strong, dominant, and weak energy conditions.   
Therefore our scalar field cannot couple with the ordinary matter.  
\item[2.]
In our model the effective gravitational constant changes with time.  
It reduces toward the past and becomes almost half of the present value at the redshift about ten.  
Because the luminosity of the ordinary star highly depends on the value of the gravitational constant ($L \propto G^{7-8}$), we expect that the detection of the ordinary stars will become much more difficult in such redshift region.  
\item[3.]
Naive extrapolation of the temporal change of the gravitational constant affects the expansion speed of the Universe at the epoch of nucleosynthesis.  
In order not to destroy the present scenario of nucleosynthesis, we need some amount of fine tuning of the parameters.  
\end{enumerate}

\par
In this article we have used the hypothetical scalar field with tiny mass and have not identified the field.  
There is a possibility that this field is the ordinary photon field with a tiny mass\footnote{In this case, we also have to reconsider the energy conditions and examine the violation of them.  }.  
Experimentally, the upper limit of photon mass is given as $6 \times 10^{-16}{\rm eV}$ from the measurement of the Jupiter's magnetic field by Pioneer-10 satellite \cite{davis75}.  
For example, the Proca field action is given by 
\be
S=\int d^4 x \sqrt{-g}\left[   
		-{1 \over 4}F_{\mu\nu}F^{\mu\nu}
		+{1 \over 2}(m^2-\xi R) B_{\mu}B_{\nu}g^{\mu\nu}
		-{1 \over 16\pi G}R
					\right], 
\label{procaaction}
\ee
where $F_{\mu\nu} \equiv \partial_{\mu} B_{\nu} - \partial_{\nu} B_{\mu}$.  
The energy momentum tensor for this model becomes
\bea
T_{\mu\nu}(x)&=&{2 \over \sqrt{-g}}{\delta S \over \delta g^{\mu\nu}} \nn\\
&=&{1 \over 4}g_{\mu\nu}F^{\alpha\beta}F_{\alpha\beta}-
		F^{\alpha}_{~\mu}F_{\alpha\nu}+
		m^2 (B_{\mu}B_{\nu}-{1 \over 2}g_{\mu\nu}B^2) \nn \\
		& &-\xi (R_{\mu\nu}-{1 \over 2}g_{\mu\nu}R)B^2 
		-\xi B_{\mu}B_{\nu}R
		-\xi  g_{\mu\nu} \Box B^2 + \xi (B^2)_{;\mu\nu}.  
\label{procaemtensor}
\eea
If we consider the simplest uniform electromagnetic field, the Universe becomes anisotropic and we should consider Bianchi type or Kantowski-Sachs type cosmologies.
The full study on this issue will be reported by the present authors in the near future.

\newpage
{\Large  Figure Captions}

\begin{itemize}
\item[Fig.1:] 
The scale factor vs. the cosmic time in the mild inflationary model. 
The scale factor is measured in units of $a_0=a(t_0)$.  
The horizontal axes is $t-t_0$ measured in units of $H_0^{-1}$.  
We took our parameters as 
$ m=100 H_0, ~~k=0, ~~\xi=-80, ~~\Omega_{\rm matter }=0.01$ and 
set the ``final'' conditions at the present time $t_0$ as
$\phi(t_0)=0.0023$.  
We notice that the piece-wise inflation is realized in the oscillatory evolution of the Universe.  
At each short inflationary phase the line becomes convex downward and this convexity elongates the cosmic age.  
The scale factor vanishes at $t=-1.45802$ where we identify the point of big bang.  

\item[Fig.2:] 
The scalar field vs. the cosmic time in the mild inflationary model.  
The scalar field is measured in units of $\sqrt{g} \equiv \sqrt{4\pi G/3}$).  
The horizontal axes is $t-t_0$ measured in units of $H_0^{-1}$.  
This graph shows highly non-linear oscillation caused by strong curvature coupling term.  

\item[Fig.3:] 
The cosmic age vs. the ``final'' condition $\phi_0$. 
The age is measured in units of $H_0^{-1}$.  
The scalar field is measured in units of $\sqrt{g} \equiv \sqrt{4\pi G/3}$. 
The value $\dot\phi_0$ is automatically fixed from the constraint equation Eq.(\ref{time-time}).
This variation of $\phi_0$ is equivalent to the variation of the identification of the present phase in the oscillating evolution.  
Each peak in the figure corresponds to the local maximum of the present expansion rate.  
The total number of H-frozen phases is different from peak to peak and is denoted in the figure.      
At the peak $\phi_0=0.0023$, this number becomes maximum.   
Note that larger number of H-frozen phase yields larger cosmic age.

\item[Fig.4:] 
The Hubble parameter vs. redshift.   
The vertical axes is normalized by $H_0$.  
Dashed line in this graph represents the expansion rate 
$H \equiv \dot a(t)/a(t)$ versus redshift $z$.  
However this is the bare Hubble parameter and is different from the observable Hubble parameter $H_{\rm obs}$ (solid line) defined in Eq.(hubble).  
It is obvious from the figure that $H_{\rm obs}$ rapidly reduces toward the past and stays almost constant with low value.  
Thin solid line is the bare Hubble parameter $H$ in the standard cosmology with $\Omega=1$.  
$H_{\rm obs}$ in this case becomes constant in $z$.

\item[Fig.5:] 
The Hubble diagram.  
The thick solid line represents the distance modulus m-M vs. Log(z) in our model. 
The thin solid line represents the same relation in the standard model with $\Omega=1$.  
The thick solid line has the same tendency as the dilute Universe (smaller $\Omega$).  

\item[Fig.6:] 
Numerically calculated $\rho$ (6a), $\rho+p$ (6b), and $\rho-|p|$ (6c) vs. cosmic time.  
The vertical axes is in units of $H_0^2 g=4 \pi G_0 H_0^2 /3$.  
The horizontal axes is $t-t_0$ measured in units of $H_0^{-1}$.  
As we can immediately see all the energy conditions are apparently violated periodically.  
Violations of the weak energy condition and dominant energy condition suggest that our hypothetical scalar field cannot couple with the ordinary matter.   

\item[Fig.7:]
The temporal change of $G_{\rm eff}(t)$ in the late-time inflationary model.  
The horizontal axes is the redshift $z$. 
The vertical axes is normalized by $G_0$.   
Within the redshift one, $G_{\rm eff}(t)$ changes up to several percent.  However it drastically changes at around redshift two and beyond.    
Because the luminosity of a star is known to be very sensitive to the value of the gravitational constant ($L \propto G^{7-8}$), distant stars would become much dimmer in our model.  

\end{itemize}
\end{document}